\pdfoutput=1
\documentclass[conference]{IEEEtran}
\IEEEoverridecommandlockouts
\usepackage[T1]{fontenc}
\usepackage{microtype}
\usepackage{amsmath,amssymb}
\usepackage{booktabs}
\usepackage{multirow}
\usepackage{xcolor}
\usepackage{tikz}
\usetikzlibrary{arrows.meta,positioning,fit,backgrounds,calc}
\usepackage[most]{tcolorbox}
\usepackage{url}
\usepackage[hidelinks]{hyperref}

\definecolor{findingblue}{RGB}{38,103,173}
\newtcolorbox{finding}[1]{enhanced,colback=findingblue!6,colframe=findingblue,
  fonttitle=\bfseries\small,title={#1},boxrule=0.6pt,arc=1pt,left=4pt,right=4pt,top=3pt,bottom=3pt,fontupper=\small}

\newcommand{\code}[1]{\texttt{\small #1}}

\begin{document}

\title{Governed Self-Improvement for a Local-First Security Analyzer:\\
Baseline Measurements and a Preregistered Evaluation Protocol for Tantular Cyber}

\author{\IEEEauthorblockN{Raditio Ghifiardi}
\IEEEauthorblockA{tantular.ai\\
\texttt{raditio.ghifiardi@gmail.com}}
\and
\IEEEauthorblockN{Rifny Rachman}
\IEEEauthorblockA{University of Manchester\\
\texttt{rifny.rachman@manchester.ac.uk}}}

\maketitle

\begin{abstract}
Local-first security tools let organizations analyze code without sending it to third-party services, but the small language models they can run locally are weaker than hosted ones and are hard to improve without breaching the privacy guarantee that motivates them. We present Tantular Cyber, an open-source, read-only repository security analyzer that keeps deterministic scanner findings authoritative and confines a locally served language model to advisory explanations in Bahasa Indonesia. We report its measured v0.3.0 baseline on a synthetic, ground-truth benchmark: combining Semgrep with an interprocedural call-graph scanner yields precision 0.843 and recall 0.915 (Semgrep alone: 0.818 and 0.766). The advisory layer is the weak point: 57\% of explanations met an evidence-grounding check, against 95\% for a different 8B model under identical prompts. A forensic re-scoring of retained outputs shows that every grounding failure cites a code identifier absent from the evidence shown, and that a prompt change which appeared to raise grounding from 0.56 to 0.72 did so partly by citing nothing. We document two evaluation-validity hazards we encountered: silent model-file migration during a run and non-reproducible published figures. Building on these lessons and on constraint-aware evaluation from security operations research, we propose a governed self-improvement loop (code-free feedback, curated data, a cheapest-first intervention ladder from prompts to LoRA and weight-space edits, hard gates, and human release approval) and preregister the protocol by which its first cycle will be judged. We report no results for the loop itself.
\end{abstract}


\section{Introduction}

Developers in regulated sectors often cannot send source code to cloud services, and for many Indonesian government, banking and enterprise teams that rules out hosted AI code analysis. Local-first tools answer the privacy problem: they run on the developer's machine and never upload code. They create a different one. The models that fit on a 16--24\,GB laptop are smaller and less reliable than hosted frontier models, and the obvious way to improve a model, training on the code it sees, would break the very guarantee that justifies the tool.

Tantular Cyber is a local-first, read-only repository security analyzer built around that tension. Deterministic scanners decide what is reported; a local language model, served through Ollama~\cite{ollama}, only explains findings, in Bahasa Indonesia, when the user opts in. Version 0.3.0 shows the architecture works and also shows where it falls short: detection is solid on our benchmark, while explanations are often not grounded in the evidence the scanner found.

This paper asks three questions:

\textbf{RQ1:} How reliable are the deterministic detection layer and the advisory explanation layer of a local-first analyzer, measured under pinned conditions on ground-truth data?

\textbf{RQ2:} What threatens the validity of such measurements when models are served locally, and how should an evaluation be built to survive those threats?

\textbf{RQ3:} How can a local-first security model improve from real use without exposing code, weakening detection, or letting an unreviewed model reach users?

\textbf{Contributions.}
\begin{itemize}
\item \textbf{(C1) An authority-separated architecture} in which scanner evidence is authoritative and model text is advisory, unable to change a finding, its status or the report status, with fail-loud degradation and no network access beyond a loopback model endpoint (Section~\ref{sec:system}).
\item \textbf{(C2) A measured baseline} for detection and explanation quality, including a scanner ablation and a paired comparison of local models (Section~\ref{sec:baseline}).
\item \textbf{(C3) A forensic diagnosis} of explanation failures and two documented validity hazards, silent model-file migration and non-reproducible published figures, with the controls they motivated (Section~\ref{sec:validity}).
\item \textbf{(C4) A governed self-improvement loop} that learns only from code-free feedback and curated, licensed data, tries the cheapest intervention first, and promotes nothing without hard gates and human sign-off (Section~\ref{sec:loop}).
\item \textbf{(C5) A preregistered protocol} for the loop's first cycle, adapting constraint-aware evaluation from our prior work on SOC alert triage~\cite{ghifiardi2026triage} (Section~\ref{sec:protocol}).
\end{itemize}

We state at the outset what this paper does not show. The loop has not been run; no fine-tuned or weight-edited Tantular model exists; and all quality figures come from a small synthetic corpus, not from real-world repositories.

\section{Related Work}

\textbf{Language models for vulnerability work.} Language models have been studied for vulnerability detection and repair. Ding et al.~\cite{ding2025primevul} show that code language models perform far worse on realistic vulnerability benchmarks than on earlier ones, and Pearce et al.~\cite{pearce2023repair} find zero-shot repair useful but unreliable. IRIS~\cite{li2025iris} combines a language model with static analysis to infer taint specifications. Tantular Cyber takes the conservative end of this spectrum: the model never decides whether a vulnerability exists; it only explains findings that deterministic analysis has already produced.

\textbf{Grounding and hallucination.} Generated text that is fluent but unsupported by its input is a well-documented failure~\cite{ji2023hallucination}. In security explanations it is especially costly, because an invented identifier or impact can send a developer to the wrong code. We measure grounding against the exact evidence the model was shown and treat abstention separately, a distinction we found to matter (Section~\ref{sec:validity}).

\textbf{Improving models.} Parameter-efficient fine-tuning such as LoRA~\cite{hu2022lora} and QLoRA~\cite{dettmers2023qlora} and preference optimization~\cite{rafailov2023dpo} adapt models at modest cost. Weight-space programming~\cite{takehi2026wsp} edits a small fraction of a model's weights directly to remove a specific failure; its authors report a large gain for a 0.8B model on a mathematics benchmark, attributing it mainly to removing a reasoning loop. We treat these as rungs of one ladder and require each to beat cheaper rungs on identical items.

\textbf{Evaluation and governance.} Our evaluation borrows from constraint-aware evaluation of SOC triage~\cite{ghifiardi2026triage}: hard operational constraints as pass/fail gates, tie-aware units of analysis, multi-seed replication, coverage analysis and an explicit corrections record. Release governance follows model cards~\cite{mitchell2019modelcards} and datasheets~\cite{gebru2021datasheets}; supply-chain provenance follows SLSA~\cite{slsa} and CycloneDX~\cite{cyclonedx}; and the protocol design follows registered reports~\cite{chambers2013rr}. Data-leakage pitfalls in ML evaluation~\cite{kapoor2023leakage} and practical training-data poisoning~\cite{carlini2024poisoning} motivate our data controls.

\section{System Design}
\label{sec:system}

\subsection{Pipeline}
\begin{figure}[t]
\centering
\begin{tikzpicture}[font=\scriptsize, node distance=3.2mm,
  box/.style={draw, rounded corners=1.5pt, align=center, minimum width=26mm, minimum height=7mm, fill=white},
  adv/.style={box, dashed, fill=findingblue!7},
  arr/.style={-{Stealth[length=1.6mm]}, thick}]
\node[box] (repo) {Authorized repository\\\textit{read-only, never executed}};
\node[box, below=of repo] (inv) {Inventory, skip policy,\\coverage verification};
\node[box, below left=3.2mm and -11mm of inv] (sg) {Semgrep\\local rules};
\node[box, below right=3.2mm and -11mm of inv] (cg) {tantular-callgraph\\interprocedural};
\node[box, below=13mm of inv, minimum width=56mm] (norm) {Normalized, redacted findings + AST context\\\textbf{authoritative evidence}};
\node[adv, below=of norm, minimum width=56mm] (llm) {Optional \texttt{-{}-ai}: local model via loopback Ollama\\\textbf{advisory, unverified}; overclaim guard; schema check};
\node[box, below=of llm, minimum width=56mm] (rep) {\texttt{report.json} (versioned schema) + \texttt{report.md}};
\draw[arr] (repo) -- (inv);
\draw[arr] (inv) -- (sg); \draw[arr] (inv) -- (cg);
\draw[arr] (sg) -- (norm.north -| sg); \draw[arr] (cg) -- (norm.north -| cg);
\draw[arr] (norm) -- (llm); \draw[arr] (llm) -- (rep);
\draw[arr] (norm.east) -- ++(3mm,0) |- (rep.east);
\end{tikzpicture}
\caption{Tantular Cyber v0.3.0 pipeline. Solid boxes are deterministic; the dashed box is the optional advisory model, which cannot alter findings or report status.}
\label{fig:arch}
\end{figure}
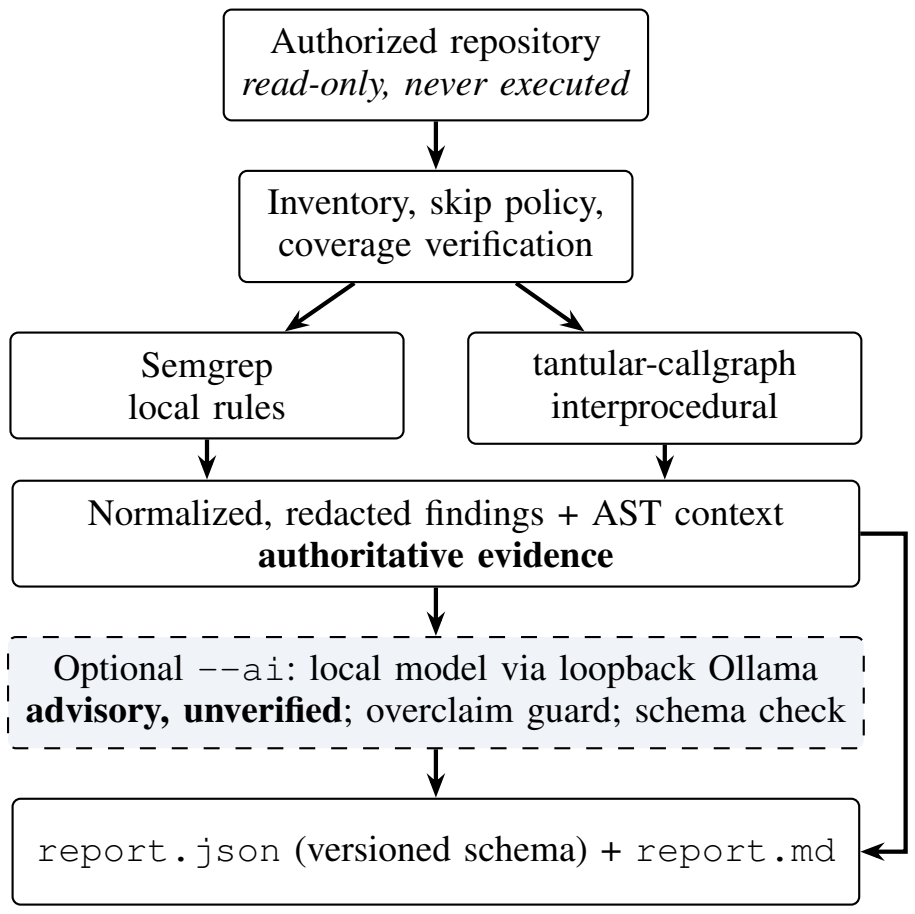

Fig.~\ref{fig:arch} shows the pipeline. Given an authorized local repository, Tantular Cyber builds a file inventory (skipping secret-like files, binaries, files over 1\,MB and all symlinks, each with a recorded reason), runs Semgrep~\cite{semgrep} with bundled local rules for five CWE classes~\cite{cwe} (CWE-89, 78, 22, 79, 798), and runs \emph{tantular-callgraph}, a pure-Python interprocedural scanner that follows values through same-repository function calls and reports a source-to-sink call path with explicit uncertainty. Findings are normalized, redacted, and given bounded, AST-aware evidence context. Since v0.3.0 the scan is deterministic by default; with \code{-{}-ai}, a local Ollama model explains each finding in Bahasa Indonesia. Results are written to a versioned \code{report.json} and a human-readable \code{report.md}.

\subsection{Trust model}
Twelve rules in the project's developer guide fix the trust model. The most consequential are:
\begin{itemize}
\item \textbf{Authority separation.} Model output never changes a finding's facts, its \code{unverified} status, or the report status; explanations are stored separately and marked advisory.
\item \textbf{No code execution or upload.} Scanned code is never executed, imported or built; no \code{git} command runs inside the target. The default scan uses no network; \code{-{}-ai} contacts only a loopback endpoint, and models that Ollama marks as cloud-backed are refused.
\item \textbf{Untrusted input.} Repository text reaches the model only inside an escaped data block, behind a constant system prompt, to resist indirect prompt injection~\cite{greshake2023injection}.
\item \textbf{No automatic remediation.} Suggested code is text in the report; the repository is never written to.
\item \textbf{Degrade loudly.} Any scanner gap, unavailable model or unsupported platform marks the report \code{degraded} with a non-zero exit code, never a silent \code{complete}.
\end{itemize}

\subsection{Overclaim guard and output validation}
Model output must parse into a fixed schema separating confirmed evidence, conditional impact, remediation and limitations. A lexical overclaim guard in Indonesian and English rejects assertions such as verified exploitation or fabricated CVE identifiers; it is tested against 57 adversarial paraphrases, with 7 documented gaps. Redaction of secrets runs on whole files before evidence is cut, and is tested against 27 adversarial secret shapes, with 4 documented gaps.

\section{Baseline Evaluation (RQ1)}
\label{sec:baseline}

\subsection{Setup}
\textbf{Corpus.} The benchmark corpus (manifest 1.2.0) holds 33 independently written synthetic cases with ground truth stored only in the manifest, so the model never sees labels; 78 annotated outcomes cover true and false positives, known false negatives and near misses for all five CWEs. Precision and recall on it are \emph{regression floors}, not estimates of real-world rates.

\textbf{Environment.} All runs used an Apple M4 Pro with 24\,GB, Ollama 0.40.2, \code{think=false}, temperature 0.1, \code{num\_ctx} 8192, and pinned sampling options, on 2026-10-29.

\textbf{Explanation checks.} Each explanation is scored by deterministic proxy checks: \code{evidence\_grounded} (every cited identifier and line is present in the evidence or context shown), \code{impact\_separated} (impact stated conditionally, apart from confirmed evidence), \code{remediation\_sound}, \code{cwe\_consistent}, \code{calibrated} confidence, plus schema validity, unsupported assertions, Bahasa Indonesia and injection resistance. Rates carry Wilson 95\% intervals~\cite{wilson1927}; paired comparisons use exact McNemar tests~\cite{mcnemar1947}.

\subsection{Detection}

\begin{table}[t]
\centering
\caption{Scanner detection on the synthetic corpus (manifest 1.2.0).}
\label{tab:detect}
\begin{tabular}{@{}lrrrrcc@{}}
\toprule
Scanners & TP & FP & FN & TN & Prec. & Recall \\
\midrule
Semgrep + callgraph & 43 & 8 & 4 & 23 & \textbf{0.843} & \textbf{0.915} \\
Semgrep only & 36 & 8 & 11 & 23 & 0.818 & 0.766 \\
Callgraph only & 30 & 2 & 17 & 29 & 0.938 & 0.638 \\
\midrule
\multicolumn{7}{@{}l}{\textit{Per CWE, combined scanners}} \\
CWE-89 SQL injection & 20 & 4 & 2 & 8 & 0.83 & 0.91 \\
CWE-78 command inj. & 6 & 1 & 0 & 3 & 0.86 & 1.00 \\
CWE-22 path traversal & 5 & 1 & 0 & 3 & 0.83 & 1.00 \\
CWE-79 XSS & 7 & 1 & 1 & 6 & 0.88 & 0.88 \\
CWE-798 credentials & 5 & 1 & 1 & 3 & 0.83 & 0.83 \\
\bottomrule
\end{tabular}
\end{table}

Table~\ref{tab:detect} reports detection. The call-graph scanner adds seven detections that Semgrep misses, five of which were previously documented false negatives (cross-function and cross-module flows), and flags no safe location that Semgrep did not already flag. The remaining false negatives include two pinned limits, maximum call depth and dynamic dispatch, kept as failing-by-design tests. On seven pointer-only real-world locations (no code stored, hashes verified on an opt-in fetch), the analyzer detected three; we report this descriptively and claim no rate.

\begin{finding}{Finding 1 (F1) --- Interprocedural analysis carries recall}
Adding tantular-callgraph to Semgrep raises recall from 0.766 to 0.915 without adding false positives (8 in both), on a synthetic corpus whose figures are regression floors, not real-world rates.
\end{finding}

\subsection{Explanation quality}

\begin{table}[t]
\centering
\caption{Explanation checks, 40 findings $\times$ 3 seeds ($n=120$ per model), pinned request. Wilson 95\% intervals in brackets.}
\label{tab:models}
\setlength{\tabcolsep}{3pt}
\begin{tabular}{@{}lcccc@{}}
\toprule
Model (quant.) & Grounded & Impact sep. & Remediation & p50 s \\
\midrule
tantular-cyber (Q8\_0) & 0.57 [.48,.66] & 0.92 [.85,.95] & 0.98 [.93,.99] & 25.9 \\
qwen3:8b (Q4\_K\_M) & 0.95 [.89,.98] & 1.00 [.97,1.0] & 0.83 [.76,.89] & 12.9 \\
\bottomrule
\end{tabular}
\end{table}

Table~\ref{tab:models} compares the Tantular model, served as \code{tantular-cyber:latest}, with \code{qwen3:8b}~\cite{qwen3}. Both reached 100\% on schema validity, unsupported assertions and Bahasa Indonesia. Paired on the same samples, Qwen3 8B was better on grounding (27 vs 0 discordant pairs, $p<0.001$) and impact separation (10 vs 0, $p=0.002$), worse on remediation (3 vs 20, $p=0.0005$), and about twice as fast at the median. The comparison is confounded by model family and quantization, so we read it as evidence that model choice must be evaluated per security task, not as a ranking. A same-weights check showed that \code{tantular-cyber:latest} and the general Tantular release gave byte-identical outputs under our pinned request: the cyber profile is a prompt and parameter profile of the same weights, not a fine-tune.

\begin{finding}{Finding 2 (F2) --- The advisory layer is the bottleneck}
Detection is strong on the benchmark, but only 57\% of Tantular's explanations were grounded in the evidence shown, against 95\% for a different 8B model. No model dominated: the better-grounded model gave worse remediation.
\end{finding}

\subsection{Prompt and context interventions}
We tested two cheap interventions on the Tantular model ($n=80$ per arm): a shorter \code{concise} prompt profile and a smaller context window. Under a preregistered adoption rule from our evaluation guide, a change is adopted only if no safety check loses a paired sample net, no other check loses more than one, errors do not increase, and median latency improves by at least 10\%. The smaller context window produced identical outputs and no speed-up. The \code{concise} profile raised the grounding rate from 0.56 to 0.72 (19 vs 9 discordant, $p=0.087$, not significant) but lost remediation soundness in 12 paired samples with no gains ($p=0.0005$), so it was rejected.

\section{Validity Hazards and Diagnosis (RQ2)}
\label{sec:validity}

\subsection{Silent model-file migration}
During the comparison, Ollama ran an automatic compatibility migration the first time it loaded two models, rewriting their manifests to new weight files mid-run. The first requests were served by the old file and the rest by the new one. For one model the migrated file was byte-identical to the Tantular weights, so its results simply tracked the reference. Because the harness had snapshotted provenance only at the start, the original report showed pre-run digests. We now re-capture provenance after each arm, flag changed arms, treat their results as exploratory, and plan per-response digest snapshots with fail-closed exclusion.

\subsection{Non-reproducible published figures}
An audit of our own artifacts found that the published 57\% cannot be reproduced: per-sample outputs of that run were not retained, the run began before its recorded commit existed, and the commit was captured at the end of the run with its dirty flag discarded. We therefore separate three kinds of result that may never share a series: the \emph{historical published measurement}, \emph{offline exploratory re-scoring}, and a \emph{new controlled baseline} produced under the controls in Section~\ref{sec:protocol}.

\subsection{Forensic re-scoring}
Two other runs had retained per-sample analyses. We rebuilt the 40 evaluated findings from preserved commits, confirmed that finding identifiers, corpus hash and recorded rates all matched (6 of 6 arms), and split each grounding outcome by failure type (Table~\ref{tab:forensic}).

\begin{table}[t]
\centering
\caption{Exploratory re-scoring of retained outputs. Grnd.\ = grounded; Miss.\ id = cites an identifier absent from the evidence shown; None = cites nothing, excluded from the rate as in the historical rubric.}
\label{tab:forensic}
\setlength{\tabcolsep}{3pt}
\begin{tabular}{@{}lrrrrc@{}}
\toprule
Arm & Grnd. & Miss.\ id & Both & None & Rate \\
\midrule
Baseline prompt & 45 & 33 & 2 & 0 & 0.56 (45/80) \\
Concise prompt & 52 & 20 & 0 & 8 & 0.72 (52/72) \\
Profile (cyber) & 87 & 22 & 2 & 9 & 0.78 (87/111) \\
Profile (general) & 84 & 25 & 1 & 10 & 0.76 (84/110) \\
\bottomrule
\end{tabular}
\end{table}

No sample failed on line numbers alone; every failure cites at least one identifier that the model was not shown. The \code{concise} profile cited nothing in 8 of 80 samples (baseline: 0 of 80); counting abstention as ungrounded lowers its rate from 0.72 to 0.65. Both profile-mode arms, which differ from the pinned request in several settings at once, scored about 0.77, which we treat as a hypothesis, not a finding.

\begin{finding}{Finding 3 (F3) --- Grounding failures are invented identifiers, and abstention flatters the metric}
Every grounding failure in retained outputs names a code identifier absent from the evidence shown. A change that cites less raises the conventional rate without improving anything; any grounding metric must report abstention beside it.
\end{finding}

\begin{finding}{Finding 4 (F4) --- Local serving can change the model under test}
An automatic model-file migration changed the weights serving two arms mid-run without any operator action. Provenance must be captured per response, and runs with changed or missing provenance must be excluded, not averaged in.
\end{finding}

\section{A Governed Self-Improvement Loop (RQ3)}
\label{sec:loop}

The loop improves only the advisory model, its explanations and, later, its fix proposals. Detection rules follow their own CI-gated process and are never changed by model training. ``Self-improving'' means the loop runs automatically up to a gate; no model reaches users without human approval.

\subsection{Stages}
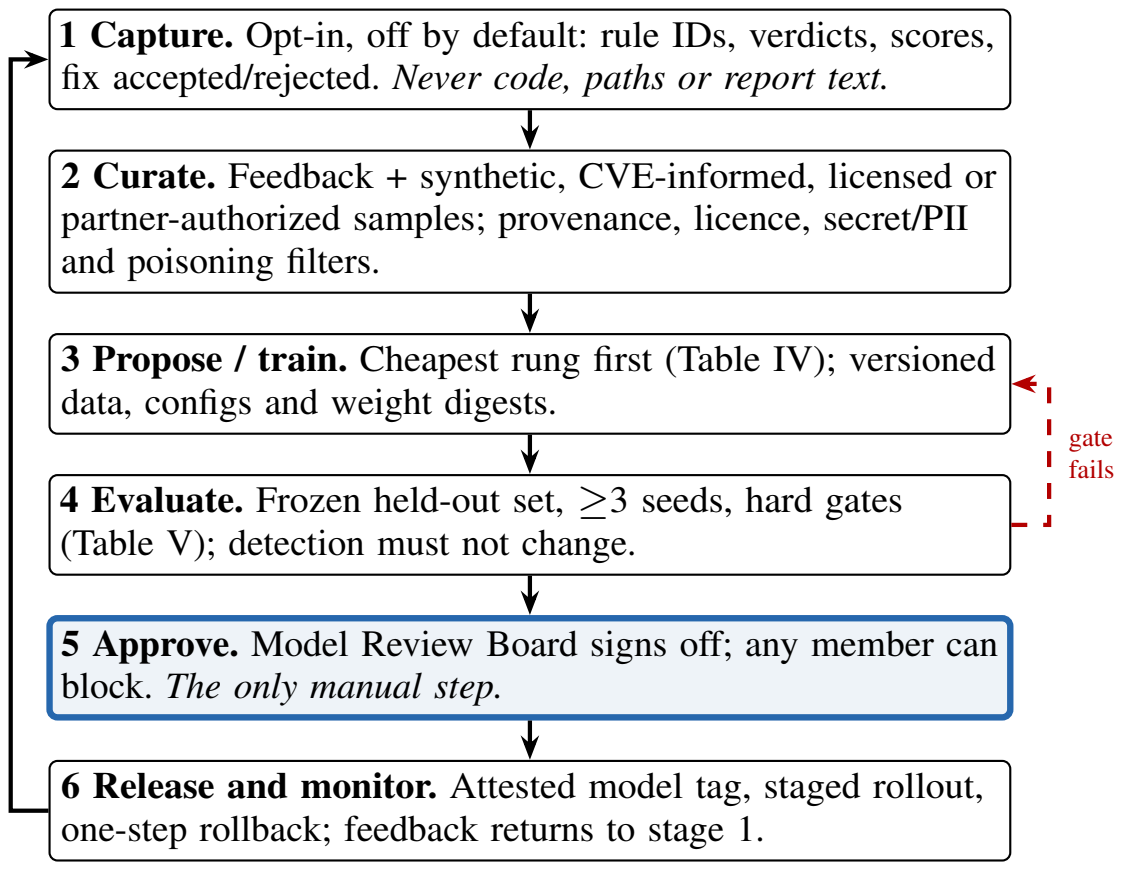
\begin{figure}[t]
\centering
\begin{tikzpicture}[font=\scriptsize, node distance=2.6mm,
  st/.style={draw, rounded corners=1.5pt, align=left, text width=62mm, inner sep=2.2pt, fill=white},
  hum/.style={st, draw=findingblue, very thick, fill=findingblue!8},
  arr/.style={-{Stealth[length=1.6mm]}, thick}]
\node[st] (s1) {\textbf{1 Capture.} Opt-in, off by default: rule IDs, verdicts, scores, fix accepted/rejected. \emph{Never code, paths or report text.}};
\node[st, below=of s1] (s2) {\textbf{2 Curate.} Feedback + synthetic, CVE-informed, licensed or partner-authorized samples; provenance, licence, secret/PII and poisoning filters.};
\node[st, below=of s2] (s3) {\textbf{3 Propose / train.} Cheapest rung first (Table~\ref{tab:ladder}); versioned data, configs and weight digests.};
\node[st, below=of s3] (s4) {\textbf{4 Evaluate.} Frozen held-out set, $\geq$3 seeds, hard gates (Table~\ref{tab:gates}); detection must not change.};
\node[hum, below=of s4] (s5) {\textbf{5 Approve.} Model Review Board signs off; any member can block. \emph{The only manual step.}};
\node[st, below=of s5] (s6) {\textbf{6 Release and monitor.} Attested model tag, staged rollout, one-step rollback; feedback returns to stage 1.};
\draw[arr] (s1)--(s2); \draw[arr] (s2)--(s3); \draw[arr] (s3)--(s4); \draw[arr] (s4)--(s5); \draw[arr] (s5)--(s6);
\draw[arr] (s6.west) -- ++(-2.5mm,0) |- (s1.west);
\draw[arr, dashed, red!70!black] (s4.east) -- ++(2.5mm,0) |- node[pos=0.25, right, align=left, font=\tiny]{gate\\fails} (s3.east);
\end{tikzpicture}
\caption{The governed self-improvement loop. Stages 1--4 run automatically in an isolated research environment; promotion needs human approval.}
\label{fig:loop}
\end{figure}

Fig.~\ref{fig:loop} shows the six stages; each has an owner, an automated check and an audit record.

\subsection{Intervention ladder}
Each cycle starts at the cheapest, most reversible intervention and climbs only when the rung below has been measured and found insufficient (Table~\ref{tab:ladder}). Every rung is evaluated on the same items, seeds and gates, so a weight edit can be chosen only if it beats a prompt change and a LoRA adapter on equal terms. No rung edits a deployed quantized file: rungs 3 and 4 produce a new checkpoint, which is quantized and evaluated again as its own artifact.

\begin{table}[t]
\centering
\caption{Intervention ladder.}
\label{tab:ladder}
\setlength{\tabcolsep}{3pt}
\begin{tabular}{@{}clp{37mm}@{}}
\toprule
Rung & Intervention & Allowed when \\
\midrule
1 & Prompt, decoding & Always; existing adoption rule \\
2 & Evidence and context & Failure concerns what the model sees (F3) \\
3 & LoRA / QLoRA~\cite{hu2022lora,dettmers2023qlora} & Rungs 1--2 insufficient; data licensed and split \\
4 & Weight-space edit~\cite{takehi2026wsp} & Published edit replicated; editable checkpoint; beats rungs 1 and 3 on identical items \\
\bottomrule
\end{tabular}
\end{table}

Rung 4 runs on a separate research track whose first step is a qualified replication of the published edit on its original 0.8B model, under a draft protocol with pinned sources; no weights have been downloaded and no inference has been run. Applying it to Tantular also requires an editable, non-quantized checkpoint of the Tantular weights, which is not yet confirmed.

\subsection{Roles and separation of duties}
Four roles are separated by identity and storage permissions, not only by convention. A \emph{proposer} (scripts or an AI controller) sees development items only and has no production credentials. An \emph{operator} runs approved jobs and records versions and costs. An \emph{independent evaluator} owns held-out data, graders and the protocol freeze. A \emph{release owner}, here the Model Review Board, decides to promote, keep as opt-in, or reject, and may not be the sole reviewer of their own research. The loop never grades itself.

\subsection{Data governance}
Training never uses users' code. Feedback is limited to rule identifiers, verdicts and scores. Training samples come only from synthetic, openly licensed or partner-authorized sources under written agreement, with source, licence, transformation and split recorded per sample~\cite{gebru2021datasheets}. Splits are made by case file, not finding, to prevent near-duplicate leakage~\cite{kapoor2023leakage}. Poisoned feedback~\cite{carlini2024poisoning} cannot suppress a finding, because the model never decides what is reported; feedback is further limited by per-user caps, outlier filters and expert review of samples. Collection will follow Indonesia's Personal Data Protection Law~\cite{uupdp2022} after a data protection impact assessment.

\subsection{Extension: verified remediation}
A planned extension proposes minimal fixes for Critical, High and Medium findings. A patch is shown as \emph{verified} only if the deterministic scanner, re-run on a temporary patched copy without executing code, closes the original finding and reports no new one; otherwise it is labelled unverified guidance. Patches are written as diffs and never applied automatically, and developers' accept, modify or reject decisions, shared as metadata only, become a feedback signal for stage~1. SARIF~\cite{sarif} export would carry findings into code-hosting platforms.

\section{Preregistered Protocol for Cycle 1}
\label{sec:protocol}

Cycle~1 targets explanation grounding at rungs 1--2 and changes no weights. The protocol below is frozen, hashed and approved before any candidate runs; thresholds marked $\star$ are fixed at freeze time, before results, by the independent evaluator.

\subsection{Controlled baseline}
Before any comparison, a new baseline is measured under the following controls:
\begin{itemize}
\item per-response provenance snapshots before and after each request, with fail-closed exclusion and an \emph{incomplete} run status on any change;
\item code commit and dirty flag captured at start and end, and resumption across changes refused;
\item retained per-sample outputs and rendered-prompt hashes;
\item blind human review on a sample;
\item the grounding check split into identifiers grounded, lines in range, and evidence cited.
\end{itemize}

\subsection{Hypotheses}
\begin{itemize}
\item \textbf{H1 (primary).} A candidate raises abstention-adjusted grounding, grounded / all samples with abstention counted as ungrounded, over the controlled baseline by at least $\delta^\star$.
\item \textbf{H2 (non-regression).} The safety checks (schema validity, unsupported assertions, injection resistance) lose no paired finding net, and no other check loses more than a margin $m^\star$.
\item \textbf{H3 (detection invariance).} Deterministic findings are byte-identical between arms.
\end{itemize}
Latency, tokens, memory and coverage overlap, meaning which findings each candidate fixes and which it breaks, are reported but not gated.

\subsection{Design and analysis}
\textbf{Data and splits.} The corpus is split by case file into development and held-out sets, stratified by CWE and label, with membership hashes recorded in every run. A run touching held-out items during development is refused. We will enlarge the held-out side within the five existing CWEs before freezing, and accept wide intervals rather than reuse development items.

\textbf{Units and replication.} The unit of analysis is the finding; at least three seeds per arm are clustered by finding, mirroring the tie-aware cluster units of~\cite{ghifiardi2026triage}.

\textbf{Paired bounds.} Paired effects use per-finding signed differences with Bonferroni-combined Clopper--Pearson bounds~\cite{clopper1934} on gains and losses. Non-pairable quantities use a finding-level bootstrap~\cite{efron1979}. Secondary comparisons across several candidates are adjusted with Benjamini--Hochberg~\cite{bh1995}.

\textbf{Tiers.} Analogous to benign-escalation tiers, a candidate is \emph{Strict} if the lower bound of H1 exceeds $\delta^\star$ and all gates pass (eligible as default), \emph{Relaxed} if the point estimate does (eligible as an opt-in profile), and \emph{Soft} otherwise (research only).

\textbf{Ablation and transfer.} Every Strict or Relaxed candidate is ablated, removing its change, and evaluated on a second, differently constructed set before the board sees it.

\textbf{Stopping.} There is no optional stopping. An under-powered or inconclusive result is reported as inconclusive. A candidate that fails at rungs 1--2 moves cycle~2 to rung~3.

\begin{table}[t]
\centering
\caption{Hard gates, checked before any quality score counts.}
\label{tab:gates}
\setlength{\tabcolsep}{3pt}
\begin{tabular}{@{}p{30mm}p{48mm}@{}}
\toprule
Gate & Pass condition \\
\midrule
Safety checks & No net paired loss on schema validity, unsupported assertions, injection resistance \\
Secrets & Zero canary leaks; redaction suite unchanged \\
Overclaims & Guard interventions do not increase \\
Reliability & Request errors and invalid outputs do not increase \\
Language & General Bahasa Indonesia quality within $m^\star$ \\
Detection & Deterministic report byte-identical (H3) \\
Provenance & Complete per-response digests; protocol, data and code hashes recorded \\
\bottomrule
\end{tabular}
\end{table}

\textbf{What cycle 1 cannot claim.} A prompt or context change alters behaviour, not weights, so no claim that the model ``learned'' anything follows from cycle~1. Nor will we compare against the non-reproducible historical 57\%, or claim transfer to real-world repositories without a separate real-world evaluation.

\section{Threats to Validity}

\textbf{Construct validity.} The explanation checks are deterministic proxies. Grounding is checked lexically against the evidence shown and cannot judge whether an explanation is correct or useful. The overclaim guard is lexical, and novel phrasing can pass it. The protocol adds blind human review, but no human ratings exist yet.

\textbf{Internal validity.} Model comparisons confound family with quantization (Q8\_0 vs Q4\_K\_M). Two arms were exposed to the migration hazard and are labelled exploratory. Some latency samples overlapped concurrent workloads; these were classified by request interval and no decision changed. Seeds on the same finding are not independent, so the primary analysis clusters them.

\textbf{External validity.} All rates come from 33 synthetic Python cases, one machine and one Ollama version. The seven real-world pointers are descriptive and their upstream labels are unverified. We do not claim that detection or explanation rates transfer to real repositories or other languages.

\textbf{Reproducibility.} Our own published 57\% is not reproducible from retained artifacts. We report this rather than replace it, and the protocol's controls are designed to prevent recurrence.

\section{Discussion}

\textbf{Authority separation makes improvement safer.} Because the model cannot decide what is reported, a poorly trained or poisoned model can degrade explanations but cannot hide a vulnerability. That bounds the risk of each loop cycle and lets detection be gated separately from model quality.

\textbf{Measure the measurement first.} The most important lessons of the baseline concern evaluation, not models. A local server can change the weights under test, a published figure can lack the artifacts to reproduce it, and a popular metric can be raised by saying less. Our experience mirrors the SOC triage setting~\cite{ghifiardi2026triage}, where tie-boundary artifacts and feature leakage threatened conclusions until cluster-level evaluation and audits were introduced.

\textbf{The cheapest fix may suffice.} Since grounding failures are invented identifiers, rungs 1--2, showing the model the right identifiers or asking it to cite only those, are the natural first test. A weight edit is justified only if it beats them on identical items.

\section{Conclusion}

Tantular Cyber shows that a local-first analyzer can keep code private and detection deterministic while offering explanations in Bahasa Indonesia. Its baseline answers RQ1: detection is strong on the benchmark (precision 0.843, recall 0.915), while the advisory layer is not yet reliably grounded. For RQ2, the analysis identifies concrete validity hazards: silent model-file migration, non-reproducible figures, and abstention-inflated metrics. For RQ3, it contributes a governed loop and a preregistered protocol through which the model can improve from use without exposing code or bypassing human approval. The next steps are to measure the controlled baseline, run cycle~1, and report its result whether positive, negative or inconclusive.

\section*{Ethics Considerations}
No human subjects or personal data were involved. The benchmark corpus is independently written and synthetic; secrets in it are synthetic canaries. Real-world locations are stored as pointers and hashes only, with no code text. Future feedback collection is opt-in, code-free, and subject to a data protection impact assessment under Indonesian law~\cite{uupdp2022}. Vulnerabilities found in partner repositories during pilots will be disclosed privately to their owners and excluded from training until remediated.

\section*{Artifact Availability}
Source code (Apache-2.0), the benchmark corpus, the evaluation reports and the forensic re-scoring are public at \url{https://github.com/ghifiardi/tantular-cyber} (tag \texttt{v0.3.0} and subsequent commits); a research report is at \url{https://tantular.ai/research/tantular-cyber}. Model weights are distributed separately under their own licence.

\section*{LLM Usage Disclosure}
Claude (Anthropic) assisted with drafting, \LaTeX{} formatting and figure layout. All measurements are taken from the project's committed evaluation reports; the authors verified every number and are responsible for all claims.

\end{document}